\documentclass[aps,prd,amsfonts,amsmath,twocolumn]{revtex4}

\usepackage{amssymb}
\usepackage{color}
\usepackage{graphics,epsfig}

\usepackage{epstopdf}

\newcommand {\be} {\begin{equation}}
\newcommand {\ba} {\begin{eqnarray}}
\newcommand {\ee} {\end{equation}}
\newcommand {\ea} {\end{eqnarray}}

\begin{document}

\title{Gravitational Form Factors in the Axial Sector from an AdS/QCD Model}

\author{Zainul Abidin and Carl E.\ Carlson}
\affiliation{
Department of Physics, College of William and Mary, Williamsburg, VA 23187, USA}

\date{April 1, 2008}

\begin{abstract}
We calculate the stress tensor, or energy-momentum tensor, form factors of the pion and of axial vector mesons in the chiral limit of a hard wall AdS/CFT model of QCD.  One (of the two) pion gravitational form factors is directly related to the second moment of the pion generalized parton distribution, thus providing a sum rule for the latter.  As was also the case for vector mesons, both the pion and the axial vector mesons appear strikingly more compact measured by the gravitational form factor than by the electromagnetic form factor.
\end{abstract}

\maketitle

%
\section{Introduction}
%

In this paper we calculate gravitational form factors, which are form factors of the stress or energy-momentum tensor,  in the axial sector using a hard-wall model of AdS/QCD.

The gauge/gravity or AdS/CFT correspondence is studied because it offers the possibility of relating nonperturbative quantities in theories akin to QCD in 4 dimensions to weakly coupled 5-dimensional gravitational theories~\cite{Maldacena:1997re,Gubser:1998bc}.  Some applications that particularly involve mesons are found in~\cite{Polchinski:2001tt,Sakai:2004cn,Hirn:2005nr,Karch:2002sh,Erlich:2005qh,Da Rold:2005zs,deTeramond:2005su,Brodsky:2006uqa,Karch:2006pv,ArkaniHamed:2000ds,Grigoryan:2007vg,Grigoryan:2007my,BoschiFilho:2005yh,Schafer:2007qy,Evans:2007sf,Erdmenger:2007cm} and other works cited therein.  The mesons studied are mainly vector and scalar mesons and topics studied include masses, decay constants, coupling constants, and electromagnetic form factors.  Less studied to date are parton distributions, be they ordinary ones, or transverse momentum dependent ones, or generalized parton distributions (GPDs).

The present authors~\cite{Abidin:2008ku} have studied gravitational form factors and the connection to GPDs for vector mesons, obtaining sum rules for the GPDs and finding that the vector meson radius appeared notably smaller when measured from a gravitational form factor than from the electromagnetic form factor.  We wish to obtain the corresponding results for the pion, a particle for which it may be easier to obtain experimental information about the GPD~\cite{Polyakov:1999gs} and for which there is already information on the electromagnetic form factor~\cite{Horn:2006tm}.

Technically, studying the pion is more involved than studying the vector mesons because there are additional terms in the action involving the chiral fields.  The ground has been broken by workers who have studied the pion electromagnetic form factor~\cite{Brodsky:2007hb,Kwee:2007dd,Grigoryan:2007wn}.    Our work is similar in its basic approach to the latter two references, but we have attempted to make the present paper reasonably self-contained.  Also, as we are studying the axial sector to learn about the pion, it requires only a small extra effort to also study the axial vector mesons $a_1$, and we quote results for these in the body of the paper.  We have limited ourselves to the chiral limit, where one can obtain analytic results for many of the quantities of interest.

In general, in AdS/CFT there is a correspondence between 4-dimensional operators $\mathcal O(x)$ and fields in the 5-dimensional bulk $\phi(x,z)$, where $z$ is the fifth coordinate.  The 4D sources used in the 4D generating function $Z_{4D}$ we will call $\phi^0(x)$, and
\be
Z_{4D}[\phi^0] = \left\langle \exp
	\Big( iS_{4D} + i \int d^4x \ {\mathcal O(x)} \phi^0(x) \Big) \right\rangle	.
\ee
The correspondence may be written as
\be
Z_{4D}[\phi^0] = e^{iS_{5D}[\phi_{cl}]}	,
\ee
where on the right, $S[\phi_{cl}]$ is the classical action evaluated for classical solutions $\phi_{cl}$ to the field equations with boundary condition
\be
\lim_{z \to 0} \phi_{cl}(x,z) = z^\Delta \phi^0(x)		\,.
\ee
The constant $\Delta$ depends on the nature of the operator $\mathcal O$, and is zero in simple cases~\cite{Son:2007vk}.

The original correspondence~\cite{Maldacena:1997re} related a strongly-coupled, large $N_c$, 4D conformal field theory  to  a weakly-coupled gravity theory on 5D AdS space. In QCD, $N_c$ is not large, nor is the theory conformal, as evidenced by the existence of hadrons with definite mass. Nonetheless, results obtained treating $N_c$ as large work surprisingly well, and one can argue that QCD behaves approximately conformally over wide regions of $Q^2$~\cite{Brodsky:2006uqa}.   We simulate the breaking of conformal symmetry,  following the so-called ``bottom-up'' approach as implemented in~\cite{Erlich:2005qh,Da Rold:2005zs},  by introducing a sharp cutoff in AdS space at $z = z_0$.   The unperturbed AdS space metric is
\be
ds^2= g_{MN} dx^M dx^N
	= \frac{1}{z^2} \eta_{MN} dx^M dx^N,	\quad \varepsilon<z<z_0,
\ee
where $\eta_{MN}=\text{diag}(1,-1,-1,-1,-1)$.  The $z= \varepsilon$ wall, with $\varepsilon \to 0$ understood, corresponds to the UV limit of QCD, and the wall located at $z=z_0 \equiv 1/\Lambda_{\rm QCD}$ sets the scale for the breaking of conformal symmetry of QCD in the IR region.   [Lower case Greek indices will run from $0$ to $3$, and upper case Latin indices will run over $0,1,2,3,5$.]


The $\mathcal O$ to $\phi$ operator correspondences of particular interest here are~\cite{Erlich:2005qh,Da Rold:2005zs,Son:2007vk}
\ba
{J_5^a}^\mu(x) &\leftrightarrow& {A^a}^\mu(x,z) , \nonumber\\
T_{\mu\nu}(x)  &\leftrightarrow& h_{\mu\nu}(x,z) ,
\ea
where ${J_5^a}^\mu=\bar{q} \gamma^\mu \gamma_5 t^a q$ is an axial current with flavor index $a$,  $A^{a\mu}$ is an axial field,  $T_{\mu\nu}$ is the stress tensor, and $h_{\mu\nu}$ represents variations of the metric tensor,
\be
g_{\mu\nu}(x,z) = \frac{1}{z^2} \left( \eta_{\mu\nu} + h_{\mu\nu}(x,z) \right)\label{AdSmetric}
\ee

We will use $h_{\mu\nu}$ in the Randall-Sundrum gauge~\cite{Garriga:1999yh}, wherein $h_{\mu\nu}$ is transverse and traceless (TT) and also satisfies $h_{\mu z} = h_{zz} = 0$.  Variations of the metric tensor in a TT gauge will only give us the transverse-traceless part of the stress tensor.  This will determine uniquely one of the two gravitational form factors of the pion, the one that enters the momentum sum rule, and correspondingly 4 of the 6 gravitational form factors for spin-1 particles, including the two that enter the momentum and angular momentum sum rules.

Relevant details regarding the pion and axial-vector mesons, including the wave functions and the two-point functions, are worked out in Sec.~\ref{sec:pion}, and Sec.~\ref{sec:ff} works out the three-point functions, and extracts from them the stress tensor matrix elements.  Sum rules and stress tensor form factor radii are given in Sec.~\ref{sec:gpd} and some conclusions are offered in Sec.~\ref{sec:theend}.

%
\section{Pion and Axial-Vector Meson} \label{sec:pion}
%
\subsection{AdS/QCD Model}
The action on the 5-dimensional AdS space is \cite{Erlich:2005qh}
\ba
S_{5D}&=&\int d^5 x \sqrt{g}\bigg\{ \mathcal{R}+12			 \label{fullAction}	\\
&&+\text{Tr}\Big[  |DX|^2+3|X|^2-\frac{1}{4g_5^2}(F_L^2+F_R^2) \Big] \bigg\}   .
					\nonumber
\ea
This action contains the $X$ field, which corresponds to 4D operator $\bar{q}_R q_L$ and, through $F^{MN}_{L,R}=\partial^M A^N_{L,R}-\partial^N A^M_{L,R}-i[A^M_{L,R},A^N_{L,R}]$, also contains the ${A^a_L}_\mu$ and ${A^a_R}_\mu$ fields, which correspond to operators ${J^a_{L}}_\mu=\bar{q}_{L}\gamma_\mu t^a q_{L}$ and ${J^a_{R}}_\mu=\bar{q}_{R}\gamma_\mu t^a q_{R}$ respectively. We define $A_M(x,z)=A^a_M(x,z) t^a$, where the group generators satisfy Tr$(t^a t^b)=\delta^{ab}/2$.
The covariant derivative of the $X$ field is given by $D^M X=\partial^M X-i A_L^M X+iX A_R^M$. Moreover, the $X$ field can be written in exponential form as $X(x,z)=X_0(z)$ exp$(2i t^a \pi^a)$. Solving the equation of motion of $X_0(z)$, one obtains $X_0=\frac{1}{2}\mathbb{I} v(z)$, where $v(z)=m_q z+\sigma z^3$. Using the AdS/CFT prescription and the fact that $\bar{q}_R q_L$ appears in the mass term of QCD Lagrangian, parameter $m_q$ can be identified as the quark mass and parameter $\sigma$ as the quark condensate $\left<\bar q q\right>$. In this paper, we will discuss only the chiral limit of the AdS/QCD model, {\it i.e} $m_\pi=0$ case, or equivalently $m_q=0$.

The axial-vector and pseudoscalar sector of the action up to second order is given by \cite{Erlich:2005qh}
\ba
S_A &=& \int d^5 x \sqrt{g} \,  \bigg[ \frac{v(z)^2}{2} g^{MN}
	(\partial_M\pi^a-A^a_M) (\partial_N\pi^a-A^a_N)
		\nonumber \\
&& 	\hskip 17 mm	
	-~  \frac{1}{4g_5^2} g^{KL} g^{MN}  {F^a}_{KM}F^a_{LN}   \bigg], \label{axialAction}
\ea
where $F^a_{KM}=\partial_{\raisebox{-1 pt}{$\scriptstyle K$}} A^a_M
	- \partial_{\raisebox{-1 pt}{$\scriptstyle M$}} A^a_K$, with  $A=(A_L-A_R)/2$.


\subsection{Equations of Motion}
Using the unperturbed 5-dimensional AdS space metric, and taking the variation over $A^a_M$ of equation (\ref{axialAction}), one obtains the equations of motion, which are expressed in 4D momentum space as
\ba
\partial_z \left(\frac{1}{z}\partial_z {A^a_\nu}_\perp\right)+\frac{q^2}{z}{A^a_\nu}_\perp-\frac{g_5^2v^2}{z^3}{A^a_\nu}_\perp=0, \label{a1eqofmotion}\\
\partial_z \left(\frac{1}{z}\partial_z \phi^a\right) +\frac{g_5^2v^2}{z^3}\left(\pi^a-\phi^a\right)=0, \label{pion1eqofmotion}\\
-q^2\partial_z \phi^a + \frac{g_5^2v^2}{z^2}\partial_z \pi^a=0, \label{pion2eqofmotion}
\ea
where the gauge choice $A_z=0$ has been imposed. The $\phi$ field comes from the longitudinal part of  $A^a_\mu={A^a_\mu}_\perp+\partial_\mu \phi^a$.

The fields above can be written conveniently in terms of bulk-to-boundary propagators as follows
\ba
A^a_\mu(q,z)&=&{{A}^{a0}_\mu(q)}_\perp \mathcal A(q,z) + {{A}^{a0}_\mu(q)}_\shortparallel \phi(q,z),\nonumber\\
\pi^a(q,z)&=&\frac{ iq^\mu}{q^2}{{A}^{a0}_\mu(q)}_\shortparallel \pi(q,z),\nonumber
\ea
with ${{A}^{a0}_\mu(q)}_\perp$ and ${{A}^{a0}_\mu(q)}_\shortparallel$ are the Fourier transform of the source functions of the 4D axial current operators ${J^a_{A,\mu}}(x)_\perp$ and ${J^a_{A,\mu}}(x)_\shortparallel$ respectively.

The $n^{th}$ KK-mode axial-vector meson's wave function, denoted by $\psi_{A,n}(z)$, is the solution of equation (\ref{a1eqofmotion}) with $q^2=m^2_{A,n}$, and with boundary conditions $\psi(0)=0$ and $\partial_z \psi(z_0)=0$. The normalization of $\psi_{A,n}$ is identical to that of the $\rho$-meson's wave function, and is given by $\int (dz/z) \psi_{A,n}(z)^2=1$~\cite{Erlich:2005qh}.
On the other hand, the pion's wave function, denoted by $\phi(z)$ or $\pi(z)$, is the solution of the coupled differential equations, {\it i.e.,} equation (\ref{pion1eqofmotion}) and (\ref{pion2eqofmotion}), with $q^2=0$ in the limit of massless pion. Furthermore, in this limit, the boundary conditions for the pion wave functions are $\phi(0)=0$, $\pi(0)=-1$ and $\partial_z \phi(z_0)=0$.
[A careful analysis is given in ref.~\cite{Erlich:2005qh} for $m_\pi$ small. There, the UV boundary conditions for the scalar wave functions are $\phi(0)=0$ and $\pi(0)=0$. However, the function $\pi(z)$ away from $z=0$ approaches $-1$ rather quickly.  For $m_\pi \to 0$, the function $\pi(z)$ equals $-1$ in essentially the entire slice of 5D AdS space, $0<z<z_0$.]

Using Eq.~(\ref{pion2eqofmotion}) and the UV boundary conditions, one finds $\pi(z)=-1$ for all $z$. Therefore equation (\ref{pion1eqofmotion}) can be rewritten in terms of $\Psi(z)=\phi(z)-\pi (z)$ as
\be
\partial_z \left(\frac{1}{z}\partial_z \Psi\right)-\frac{g_5^2 v^2}{z^3}\Psi=0.
\ee
The solution is given by \cite{Grigoryan:2007wn}
\be
\Psi(z)=z\Gamma[2/3]\left(\frac{\alpha}{2}\right)^{\frac{1}{3}}\left(I_{-\frac{1}{3}}(\alpha z^3)-I_{\frac{1}{3}}(\alpha z^3) \frac{I_{\frac{2}{3}}(\alpha z_0^3)}{I_{-\frac{2}{3}}(\alpha z_0^3)}\right), \label{pionSol}
\ee
where $\alpha= g_5\sigma/3$, with $g_5=2\pi$ as shown in ref~\cite{Erlich:2005qh}. Note that $\Psi(z)$ is identical to $\mathcal{A}(0,z)$. Parameter $z_0=1/\Lambda_{QCD}$ is determined by the experimental value of $\rho$-meson's mass $m_\rho=775.5 {\rm\ MeV}$~\cite{Yao:2006px}, which corresponds to $z_0=1/(322 {\rm\ MeV})$~\cite{Erlich:2005qh}.

For the $a_1$'s wave functions, one has to rely on numerical methods. However, other aspects of the axial-vector mesons are analogous to the vector mesons. For instance, the bulk-to-boundary propagator of $a_1$ can be written in terms of $\psi_{A,n}$ as
\be
\mathcal{A}(q,z)=\sum_{n} \frac{(\frac{1}{\varepsilon}\partial_{z'}\psi_{A,n}(\varepsilon))\psi_{A,n}(z)}{m^2_{A,n} - q^2 },
\ee
{\it c.f}~ equation (17) of ref.~\cite{Abidin:2008ku} and ref.~\cite{Erlich:2005qh,Grigoryan:2007vg}. The bulk-to-boundary propagator $\mathcal{A}(q,z)$ satisfies $\mathcal{A}(q,\varepsilon)=1$ and $\partial_z \mathcal{A}(q,z_0)=1$.
\subsection{Two-Point Functions}

The completeness relation is given by
\be
\sum_n  \int \frac{d^3q}{(2\pi)^3 2q^0 } \left|n(q)\right>\left<n(q)\right| =1 \, . \label{completeness}
\ee
The complete set of states includes $\left|A^a_n(q,\lambda)\right>$, the $n^{th}$ axial-vector state, as well as  $\left|\pi^a(q)\right>$, the pion state.

By applying the completeness relation into $\left<0| \mathcal{T} {J^\mu_A(x)_\perp} {J^\nu_A(0)_\perp}|0\right>$ and $\left<0| \mathcal{T} {J^\mu_A(x)_\shortparallel} {J^\nu_A(0)_\shortparallel}|0\right>$, then multiplying $q^2-m_n^2$ and taking the limit $q^2\rightarrow m_n^2$, one can extract the following quantities from the AdS/QCD correspondence~\cite{Erlich:2005qh}
\ba
F_{A,n}&=& \frac{\partial_{z}\psi_{A,n}(z)}{g_5\, z}\bigg|_{z=\varepsilon},\\
f^2_\pi&=&-\frac{\partial_z\Psi(z)}{g_5^2\,z}\bigg|_{z=\epsilon},\label{pionDecay}
\ea
where $F_{A,n}$ is the decay constant of the $n^{th}$ mode of the $a_1$, and $f_\pi$ is that of the pion. The latter is obtained in the chiral limit $q^2\rightarrow m_\pi^2=0$. The decay constants are defined by
\ba
\left<0|J^a_{A,\mu}(0)_\perp|A^b_n(p,\lambda)\right>&=&F_{A,n} \varepsilon_\mu(p,\lambda)\delta^{ab},\\
\left<0|J^a_{A,\mu}(0)_\shortparallel|\pi^b(p)\right>&=&f_\pi p_\mu\delta^{ab}.
\ea

Equation (\ref{pionDecay}) and (\ref{pionSol}) relate the input parameter $\sigma$ to the pion decay constant
\be
f^2_\pi=\frac{3}{4\pi^2}\frac{\Gamma(2/3)}{\Gamma(1/3)}\left(2\alpha^2\right)^{\frac{1}{3}}
       \frac{I_{\frac{2}{3}}(\alpha z_0^3)}{I_{-\frac{2}{3}}(\alpha z_0^3)}.
\ee
Using the experimental value $f_\pi=92.4{\rm\ MeV}$, we find $\alpha=2.28~\Lambda_{QCD}^3$, therefore $\sigma=(332 {\rm\ MeV})^3$. Consequently, other observables can be determined, $m_{A,1}=1376 {\rm\ MeV}$ and $F^{1/2}_{A,1}=493 {\rm\ MeV}$. They are in a good agreement with the experimental values $m_{A,1}=1230 {\rm\ MeV}$ and $F^{1/2}_{A,1}=433 {\rm\ MeV}$.

\section{Gravitational Form Factors} \label{sec:ff}

The three-point function that includes the stress tensor follows from
\ba
&& \big< 0 \big| {\mathcal T} {J}_5^{a\alpha}(x) \hat{T}_{\mu\nu}(y){J}_5^{b\beta}(w)\big| 0 \big>	
	\nonumber \\
&& \hskip 19 mm= -
	\frac{    2 \, \delta^3 S}{\delta A^{a0}_\alpha (x) \delta h^{\mu\nu 0}(y) \delta A^{b0}_\beta(w)} ,
		\label{ttEnergyMom}
\ea
and
the relevant part of the action (\ref{fullAction}) that contributes to the 3-point function is linear in $h^{\mu\nu}$ and quadratic in the non-gravitational fields,
\ba
S_A^{(3)} &=&
	\int {d^5 x}\bigg[ - \frac{v(z)^2 h^{\rho\sigma} }{2z^3}
    (\partial_\rho\pi^a-A^a_\rho)(\partial_\sigma\pi^a-A^a_\sigma)
				\nonumber\\
    &&  + \frac{1}{2g_5^2z} h^{\rho\sigma}
		\big[-F_{\sigma z}F_{\rho z}+\eta^{\alpha\beta}F_{\sigma\alpha}F_{\rho\beta}\big]
		\bigg],
\ea
where $h^{\rho\sigma}$ is the metric perturbation defined analogously to equation (\ref{AdSmetric}), {\it viz.}, $g^{\rho\sigma} = z^2 ( \eta^{\rho\sigma} - h^{\rho\sigma} )$.

To isolate the pion-to-pion elastic stress tensor matrix elements from the Fourier transformed 3-point functions $\left<J^\alpha(-p_2) T^{\mu\nu}(q)J^\beta(p_1)\right>$, we apply the completeness relation (\ref{completeness})
twice, then multiply by
\be
p_1^\alpha p_2^\beta \frac{1}{f_\pi^2} \,,
\ee
and take the limit $p_1^2\rightarrow m_\pi^2=0$ and $p_2^2\rightarrow m_\pi^2=0$.

We obtain the transverse-traceless part of the stress tensor matrix elements, for $\hat{T}^{\mu\nu}(0)$ at the origin in coordinate space,
\ba
&&\left<\pi^a(p_2)\right|\hat{T}^{\mu\nu}(0)\left|\pi^b(p_1)\right>
	\\	\nonumber
&&\qquad = 2 \delta^{ab} A_\pi(Q^2)\bigg[ p^\mu p^\nu+\frac{1}{12}\left(q^2\eta^{\mu\nu}-q^\mu q^\nu\right)\bigg],\label{pionTracelessMatrix}
\ea
where $p=(p_1+p_2)/2$ and $q=p_2-p_1$.
The gravitational form factor $A_\pi$ is given by
\be
A_\pi(Q^2)=\int dz\, \mathcal{H}(Q,z)\left(\frac{(\partial_z \Psi(z))^2}{g_5^2f_\pi^2z}+\frac{v(z)^2\Psi(z)^2}{f_\pi^2 z^3}\right).\label{Aformfactorpi}
\ee
Note that, except for the $\mathcal{H}$, this form factor is similar to the electromagnetic form factor given in \cite{Kwee:2007dd, Grigoryan:2007wn} although they come from different terms of the action (\ref{fullAction}). $\mathcal{H}(Q,z)$ is the bulk-to-boundary propagator of the graviton for spacelike momentum transfer $q^2=-Q^2<0$. It is defined by $h_{\mu\nu}(q,z)=\mathcal{H}(Q,z) h^0_{\mu\nu}(q)$, where $h_{\mu\nu}(q,z)$ is the Fourier transform of the metric perturbation $h_{\mu\nu}(x,z)$. In transverse-traceless gauge, $q^\mu{h}_{\mu\nu}=0$ and ${h}^\mu_\mu=0$,  the linearized Einstein equation becomes
\be
z^3\partial_z\big(\frac{1}{z^3}\partial_z {h}_{\mu\nu}\big)+q^2 {h}_{\mu\nu}=0,\label{ttgraviton}
\ee
with boundary conditions $h(q,\varepsilon)=1$ and $\partial_z h(q,z_0)=0$.
The solution is given by~\cite{Abidin:2008ku}
\be
\mathcal{H}(Q,z)=\frac{1}{2}Q^2z^2\bigg(\frac{K_1(Qz_0)}{I_1(Qz_0)}I_2(Qz)+K_2(Qz)\bigg). \label{H}
\ee
Since $\mathcal{H}(0,z)=1$, one can check that $A_\pi(0)=1$, which is a correct normalization for $A_\pi$.

Our procedure obtains the transverse-traceless part of the stress tensor;  the full stress tensor can have a trace, which means there could be a term $\frac{1}{3}(\eta^{\mu\nu}-q^\mu q^\nu/q^2)T$, where $T$ is the trace of $T^{\mu\nu}$.   In general, there are two gravitational form factors for spin-$0$ particles. The expression for the pion matrix elements written in terms of the two independent form factors is
\ba
&&\left<\pi^a(p_2)|{T}^{\mu\nu}(0)|\pi^b(p_1)\right>=
	\\	\nonumber
&&\quad \delta^{ab} \bigg[2 A_\pi(Q^2) p^\mu p^\nu+\frac{1}{2}C_\pi(Q^2)\left(q^2\eta^{\mu\nu}-q^\mu 	 q^\nu\right)\bigg];
\label{pionMatrix}
\ea
we have calculated $A_\pi(Q^2)$, but $C_\pi(Q^2)=A_\pi(Q^2)/3+\widetilde{C}_\pi(Q^2)$ where $\widetilde C$ is not determined here.

For $a_1$, the corresponding matrix element is identical to the $\rho$-meson's~\cite{Abidin:2008ku}.
The only difference is that the $\rho$-meson's wave function $\psi_n$ is replaced by $\psi_{A,n}$, the $a_1$'s wave function. The $A$ form factor is now given by
\be
A_{a_1}(Q^2)=\int \frac{dz}{z} \mathcal{H}(Q,z)\psi_{A,n}\psi_{A,n}.
\ee
The other form factors mirror the $\rho$-meson's form factors expression.

\begin{figure}[ht]
\includegraphics[width = 3.35 in]{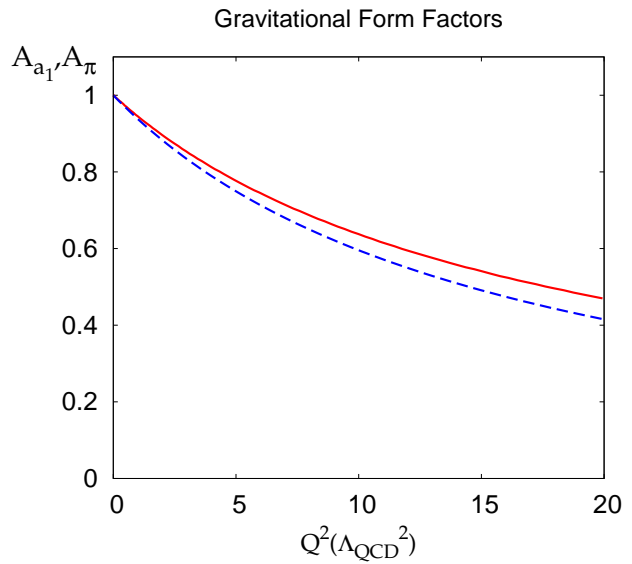}
\vglue -5 mm
\caption{Plot of $A$ with momentum transfer squared in units of $\Lambda_\text{QCD} = 1/z_0$.  The red solid line is $A_{\pi}$ and the blue dashed line is~$A_{a_1}$}
\label{fig:Aplots}
\end{figure}

Both $A_{\pi}(Q^2)$ and~$A_{a_1}(Q^2)$ are shown in Fig.~\ref{fig:Aplots}.

%
\section{Consequences} \label{sec:gpd}
%
\subsection{Radii}
In the limit $Q z_0\ll 1$, one can expand $\mathcal{H}(Q^2,z)$ and obtain the radius
\ba
\left<r^2_\pi\right>_{\rm grav}&\equiv&-6 \frac{dA_\pi}{dQ^2}\bigg|_{Q^2=0}\nonumber\\
&=&\frac{6}{4}\int dz\,z^3\left(1-\frac{z^2}{2z_0^2}\right)\rho(z),
\ea
where
\be
\rho(z)= \left(\frac{(\partial_z \Psi(z))^2}{g_5^2f_\pi^2z^2}+\frac{v(z)^2\Psi(z)^2}{f_\pi^2 z^4}\right).
\ee

\noindent We find
\be
\left<r^2_\pi\right>_{\rm grav}=0.13\, ({\rm fm})^2 = (0.36\, {\rm fm})^2 \,.
\ee
M. Polyakov, in a different model~\cite{Polyakov:1998js}, also found a small gravitational radius for the pion, $\left<r^2_\pi\right>_{\rm grav}=(0.42\, {\rm fm})^2$.  The gravitational RMS radius is significantly smaller than the electromagnetic radius $\left<r^2_\pi\right>_C= 0.33\,({\rm fm})^2 = (0.57\, {\rm fm})^2$, obtained from the AdS/QCD model in chiral limit~\cite{Grigoryan:2007wn}. Compared to the experimental result $\left<r^2_\pi\right>_C =  (0.67 \, {\rm fm})^2$~\cite{Yao:2006px}, the difference is even more apparent. This shows, as in the $\rho$-meson case, that the energy distribution of the pion is concentrated in a smaller volume than the charge distribution.

For $a_1$, the corresponding radius is
\ba
\left<r^2_{a_1}\right>_{\rm grav}&=&\frac{6}{4}\int dz\,z\left(1-\frac{z^2}{2z_0^2}\right)\psi_{A,1}^2\nonumber\\
&=&0.15 \,({\rm fm})^2 = (0.39 {\rm\ fm})^2.
\ea
As expected, it is smaller than the charge radius $\left<r^2_{a_1}\right>_C=0.39\,({\rm fm})^2 = (0.62\, {\rm fm})^2$ calculated from the AdS/QCD model.

\subsection{High $Q^2$ limit}

For high $Q^2$, the form factor $A_\pi$ scales as $1/Q^2$. The precise expression is given by
\be
A_\pi(Q^2)=\frac{4\rho(0)}{Q^2}=\frac{16\pi^2 f_\pi^2}{Q^2},
\ee
which follows from the fact that at high $Q^2$, the second term of the function $H(Q,z)$ in equation (\ref{H}) dominates. This term behaves like $e^{-Qz}$. Therefore, one can allow $z_0\to\infty$ in equation (\ref{Aformfactorpi}), and replace $\rho(z)$ by its value at the origin, $\rho(0)$, and then do the integral analytically.

One can verify, most easily in the Breit frame, that this scaling agrees with the perturbative QCD prediction. It can be shown that~\cite{Carlson:1984wr}
\be
\label{eq:highq}
\left<p_2|\eta_{\mu\nu}T^{\mu\nu}|p_1\right>\sim Q^0,\quad \left<p_2\big|e_{\mu\nu}(q,0)T^{\mu\nu}\big|p_1\right>\sim Q^0;
\ee
$e_{\mu\nu}(q,0)$ gives the helicity-$0$ component of the spin-2 part of the stress tensor, with
\ba
e_{\mu\nu}(q,0) &=& \frac{1}{\sqrt{6}}\Big( 2\zeta_\mu(q,0)\zeta_\nu(q,0)
		-  \zeta_\mu(q,+)\zeta_\nu(q,-)
					\nonumber \\
&& \hskip 8 mm   - \ \zeta_\mu(q,-)\zeta_\nu(q,+)\Big).
\ea
Here, $\zeta(q,\lambda)$ is the polarization vector of a spin-1 particle of momentum $q$.
Equation~(\ref{eq:highq}) is equivalent to
\be
A_\pi,C_\pi \sim 1/Q^2.
\ee

For $a_1$, the high $Q^2$ behavior of the form factor $A_{a_1}$ is
\be
A_{a_1}(Q^2)=\frac{12|\psi''_{A,n}|^2}{Q^4}.
\ee
Similarly $C_{a_1}, D_{a_1} \sim 1/Q^4$, while $F_{a_1}\sim 1/Q^6$, which mirror the scaling results for $\rho$-mesons, with the notation given in~\cite{Abidin:2008ku}.
\subsection{Sum Rules for the GPD}
Deeply virtual Compton scattering process involves a target absorbing a virtual photon and subsequently radiating a real photon. The virtual Compton scattering amplitude can be written in terms of integral involving the generalized parton distributions (GPD) $H(x,\xi,Q^2)$~\cite{Diehl:2003ny}. In a model with quarks,   $x$ is the light-cone momentum fraction of the struck quark constituent relative to the total momentum of the target hadron.

For spin-$0$ hadrons, there is only one GPD, defined by
\ba
&&\int \frac{p^+ dy^-}{4\pi}e^{ixp^+y^-/2}
\left<p_2\big|\bar \psi_q (-\frac{y}{2})\gamma^+
	\psi_q (\frac{y}{2})\big|p_1\right>_{y^+=y_\perp=0}
		\nonumber	\\[1.2ex]
&&\qquad=2p^+H(x,\xi,Q^2),
\ea
where $q^+= q^0 + q^3 = -2\xi p^+$.

There are sum rules connecting this GPD to the gravitational as well as to the electromagnetic form factors. The well known sum rule is for the first moment of the GPD~\cite{Ji:1996ek}
\be
\int_{-1}^{1}dx \,  H(x,\xi,Q^2)=F(Q^2),
\ee
where $F(q^2)$ is the electromagnetic form factor defined by
\be
\left<\pi(p_2)|\bar \psi_q (0)\gamma^\mu \psi_q(0)|\pi(p_1)\right>=2p^\mu F(Q^2).
\ee

A further sum rule exists because the stress tensor element $T^{++}$ can be related to the second moment in $x$ of the operator whose matrix element defines the GPDs.  For the pion, the result was given in~\cite{Polyakov:1999gs} and reads,
\be
\int_{-1}^{1} dx\, x \, H(x,\xi,Q^2) = A_\pi(Q^2)-\xi^2 C_\pi(Q^2).
\ee
Ref.~\cite{Polyakov:1999gs} also uses a chiral Lagrangian to show that $A_\pi(0)=1$ (the momentum sum rule) and $C_\pi(0) = 1/4$.    One can set $\xi=0$ in the above equation so that only the first term, which is known from the AdS/QCD model, in the right-hand side survives.  There are also first-moment sum rules for the axial vector meson GPDs, which precisely parallel the ones given for the $\rho$-mesons in~\cite{Abidin:2008ku}

%
\section{Conclusions}     \label{sec:theend}
%

We have worked out the gravitational form factors of pions and of axial-vector mesons using the AdS/CFT correspondence, and have given the sum rules connecting the gravitational form factors, which can also be called stress tensor or energy-momentum tensor form factors, to the axial sector GPDs.

A striking numerical result is the smallness of the pion radius and of the axial-vector meson radius as obtained from $A(q^2)$, the gravitational form factor that enters the momentum sum rule.  This parallels the results for the ordinary vector mesons~\cite{Abidin:2008ku}.  It suggests that the energy that makes up the mass of the meson is well concentrated, with the charge measured by the electromagnetic form factors spreading more broadly.

Extensions of the present work include considering flavor decompositions~\cite{Sakai:2004cn,Karch:2002sh,Kruczenski:2003be} of the stress tensor and the related topic of considering quarks of differing masses, to be able to separate contributions with differing internal quantum number, including strangeness.  Further, we would like work on applying the present considerations to nucleons~\cite{deTeramond:2005su}.  We hope to return to these topics, but for now they lie beyond the scope of this paper.

\begin{acknowledgments}
We thank Josh Erlich, Hovhannes Grigoryan, Herry Kwee, Anatoly Radyushkin, and Marc Vanderhaeghen for conversations and suggestions, and the National Science Foundation for support under grant PHY-0555600.
\end{acknowledgments}

\end{document}